\documentclass[11pt]{article}

\usepackage{amsmath,amsfonts,bm}

\def\eqref#1{equation~\ref{#1}}

\def\1{\bm{1}}

\DeclareMathAlphabet{\mathsfit}{\encodingdefault}{\sfdefault}{m}{sl}
\SetMathAlphabet{\mathsfit}{bold}{\encodingdefault}{\sfdefault}{bx}{n}

\usepackage[]{acl}

\usepackage{times}
\usepackage{latexsym}
\usepackage[T1]{fontenc}
\usepackage[utf8]{inputenc}
\usepackage{microtype}

\usepackage{url}
\usepackage{graphicx}  
\usepackage{amsfonts}
\usepackage{CJK}
\usepackage{bm}
\usepackage{mathrsfs}
\usepackage{float}
\usepackage{xcolor,colortbl}
\usepackage{adjustbox}
\usepackage{subfigure}
\usepackage{booktabs,multirow}
\usepackage{bigstrut}
\usepackage{paralist}
\usepackage{bbm}
\usepackage{makecell}
\usepackage{nicefrac}       
\usepackage{microtype} 
\usepackage{epsfig}
\usepackage{diagbox}
\usepackage{framed}
\usepackage{empheq}
\usepackage{array}
\usepackage{enumitem}
\usepackage{mdwlist}
\usepackage{xspace,mfirstuc,tabulary}
\usepackage{tcolorbox}
\usepackage{algorithm}
\usepackage{placeins}
\usepackage{algpseudocode}
\usepackage{microtype}

\usepackage[normalem]{ulem}
\useunder{\uline}{\ul}{}
\usepackage{threeparttable}
\usepackage{makecell}
\usepackage{booktabs}

\usepackage{xspace}
\usepackage{natbib}
\usepackage{hyperref}
\usepackage{url}

\usepackage{graphicx}
\usepackage{subfigure}
\usepackage{multirow}
\usepackage{multicol}
\usepackage{pifont}
\usepackage{enumitem}
\usepackage{amsmath}
\usepackage{appendix}
\usepackage{wasysym}

\usepackage{tikz}
\newcommand*{\circled}[1]{\lower.7ex\hbox{\tikz\draw (0pt, 0pt)%
    circle (.5em) node {\makebox[1em][c]{\small #1}};}}

\usepackage{amsmath}
\usepackage{amssymb}
\usepackage{mathtools}
\usepackage{amsthm}
\usepackage{listings} %
\newcommand{\jjx}[1]{\textcolor{red}{}}
\newcommand{\zj}[1]{\textcolor{pink}{}}

\title{Focused-DPO: Enhancing Code Generation Through Focused Preference Optimization on Error-Prone Points}

\author{Kechi Zhang$^{1,2}$, \ Ge Li$^{1,2*}$, \ Jia Li$^{3}$, Yihong Dong$^{1,2}$, Jia Li$^{1,2}$,\ Zhi Jin$^{1,2*}$ \\
$^1$Key Lab of High Confidence Software Technology (PKU), Ministry of Education \\
$^2$School of Computer Science, Peking University, China \\
$^3$College of AI, Tsinghua University \\
\texttt{\{zhangkechi,lige,zhijin\}@pku.edu.cn}}

\begin{document}
\maketitle

\renewcommand{\thefootnote}{\fnsymbol{footnote}}
\footnotetext[1]{Ge Li and Zhi Jin are the corresponding authors.}
\renewcommand{\thefootnote}{\arabic{footnote}}

\begin{abstract}

Code generation models have shown significant potential for automating programming tasks. However, the challenge of generating accurate and reliable code persists due to the highly complex and long-reasoning nature of the task.
Even state-of-the-art models often fail in code generation due to small errors, which can drastically affect the overall functionality of code.
Our study identifies that current models tend to produce errors concentrated at specific error-prone points, which significantly impacts the accuracy of the generated code.
To address this issue, we introduce Focused-DPO, a framework that enhances code generation by directing preference optimization towards these critical error-prone areas. This approach builds on Direct Preference Optimization, emphasizing accuracy in parts prone to errors. Additionally, we develop a method called Error-Point Identification, which constructs a dataset that targets these problematic points without requiring costly human annotations.
Our experiments on benchmarks such as HumanEval(+), MBPP(+), and LiveCodeBench demonstrate that Focused-DPO significantly improves the precision and reliability of code generation, reducing common errors and enhancing overall code quality. By focusing on error-prone points, Focused-DPO advances the accuracy and functionality of model-generated code.

\end{abstract}


\maketitle

\section{Introduction}

Code generation has emerged as a pivotal task in artificial intelligence, enabling models to automate essential software development tasks. 
Code Models \cite{GPT-4, guo2024deepseek,qwencoder} have demonstrated remarkable capabilities in code generation tasks.
These advancements have significantly improved developer's productivity, accelerating software delivery timelines.

Despite their success, generating correct code remains a substantial challenge due to the complex and long-reasoning nature of the task. 
Writing code necessitates long reasoning, where numerous small decisions about syntax and logic must work together to produce a functional program. 
Even minor mistakes, such as an incorrect operator, can cause a program to fail.
Code generation, therefore, can be viewed as a multi-step long reasoning process. 
Ensuring the accuracy of every decision in this multi-step process collectively determines the correctness of the resulting output code.

\begin{figure}[t]
\centering
  \includegraphics[width=\columnwidth]{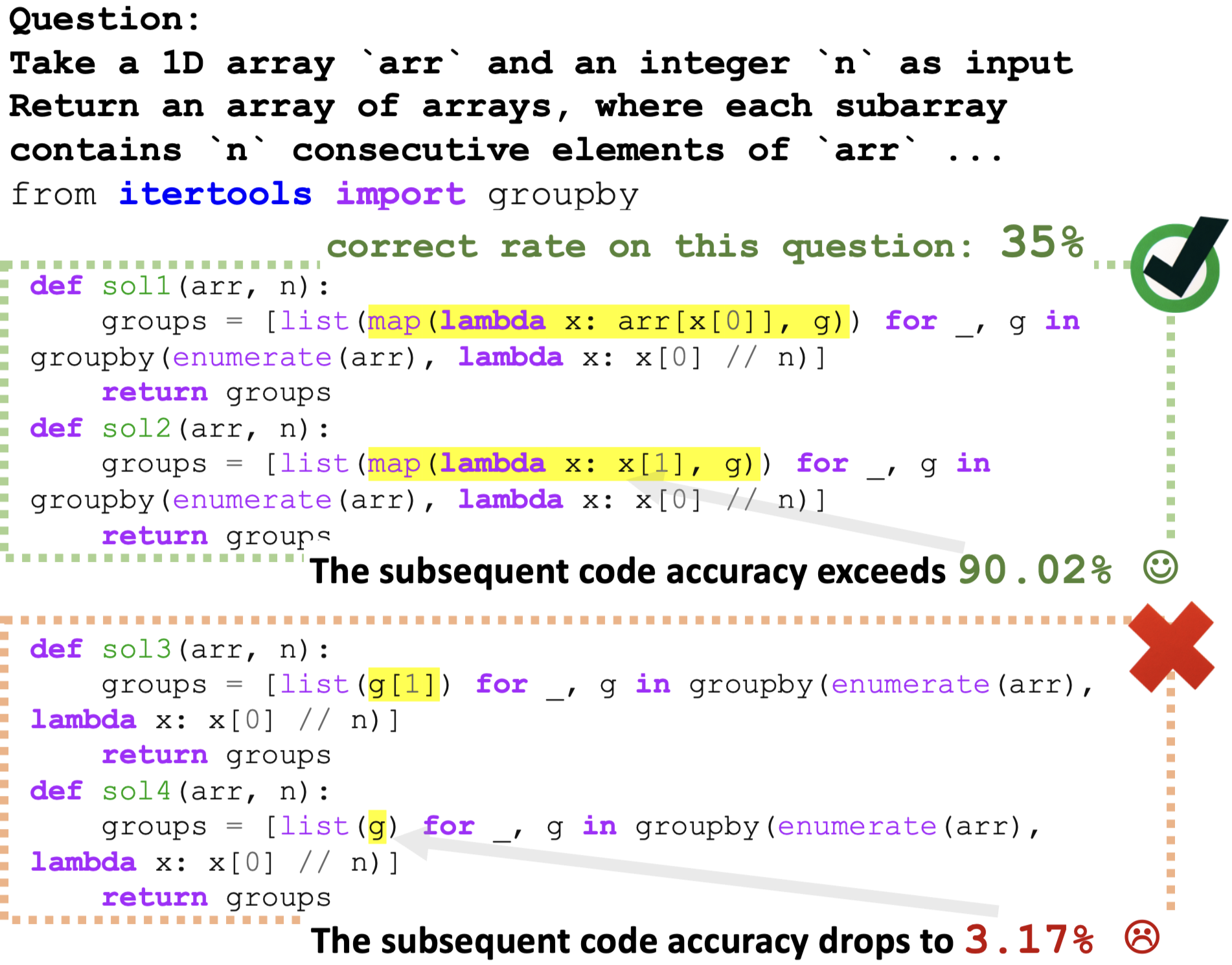}  
\caption{Error-prone points in generated code from Qwen-2.5-Coder-Instruct-7B. We sample 20 outputs for this question. Outputs have common prefixes and suffixes, differing mainly at yellow-highlighted error points. Continuing generation at these points leads to drastically different accuracies (90.02\% vs. 3.17\%). This disparity is not seen in non-highlighted parts.}
\label{fig:motivationExample}
\end{figure}

When examining the outputs of current code generation models, we find that errors are not evenly spread across the code. 
Large language models tend to produce errors concentrated in certain error-prone points, even when sampling multiple times with a high temperature.
We illustrate this phenomenon in Figure \ref{fig:motivationExample}, which shows error-prone points highlighted in yellow. 
Despite the overall code having similar prefixes and suffixes, differences at these highlighted error points significantly impact the final code accuracy. 
Generating code from correct outputs at these error-prone points can achieve a final accuracy of up to 90.02\%, whereas starting from incorrect outputs reduces accuracy to 3.17\%.
Parts of the code, such as function headers (usually at the prefix) or return statements (usually at the suffix), often follow familiar patterns. However, some middle parts of the code, which involve more complex reasoning, are more prone to errors. 
Errors in these parts can disrupt and affect the entire program's reliability.

It is crucial to address these error-prone points for code generation. However, existing studies on code generation overlook this problem.
While standard training approaches such as Supervised Fine-Tuning (SFT) \cite{wang2022self} help improve overall output quality, they do not specifically focus on the crucial parts necessary for correctness.
Methods like Direct Preference Optimization (DPO) \cite{rafailov2024direct} aim to align outputs with preferences (e.g., "chosen" vs. "rejected"), but often overlook fine-grained error-prone points of the code. 
As a result, these trained models might generate code that appears correct initially but contains critical issues at the error-prone points, ultimately affecting overall accuracy.

To tackle these issues, we introduce \textbf{Focused-DPO}, a framework designed to enhance code generation through focusing preference optimization on error-prone points. 
Focused-DPO builds on Direct Preference Optimization by emphasizing accuracy improvement in areas where errors are most likely to occur. 
Unlike traditional methods that treat all parts of the code equally, Focused-DPO specifically targets those error-prone points, which are essential for the overall correctness of the program.

Focused-DPO is a data-driven preference optimization method that relies on a specially constructed dataset with identified error-prone points. We propose a dataset construction method named \textbf{Error-Point Identification}, which includes an automated pipeline to construct paired code preference datasets. This method extracts concepts from real code repositories and synthesizes programming problems. By concurrently generating code and tests, and using a page-rank-inspired algorithm for ranking, we determine the relative performance of all generated code. Error-Point Identification employs common prefix and suffix matching to precisely locate error-prone points.
Additionally, our method automatically identifies error-prone code parts, eliminating the need for costly human input, making it scalable and efficient for a variety of programming tasks.

We evaluate Focused-DPO using standard benchmarks such as HumanEval(+) \cite{liu2024your}, MBPP(+), and LiveCodeBench \cite{jain2024livecodebench}, and observe significant improvements over existing methods. Even for models like \textit{Qwen2.5-Coder}, which already have undergone large-scale alignment training, Focused-DPO still achieves a 42.86\% relative improvement on extremely hard competition-level problems in LiveCodeBench. 
The results show notable increases in the generation quality on error-prone points, highlighting Focused-DPO's effectiveness in enhancing the accuracy of code generation.

Our contributions are summarized as follows:
\begin{itemize}
\item We propose Focused-DPO , a novel framework that enhances code generation by focusing preference optimization on error-prone points, resulting in more accurate codes.
\item We introduce a dataset construction method that automatically identifies error-prone points by generating both code and corresponding tests for fine-grained self-verification.
\item Experiments on widely-used benchmarks show that Focused-DPO improves the generation quality of code models, even for those that have already undergone extensive post-training on million-level datasets.
\end{itemize}

\section{Related Work}

Large language models (LLMs) have made significant progress in generating code from natural language descriptions, showing great potential for automating software development tasks \cite{zhang2024codeagent,zhang2024hirope}. Models\citep{GPT-4,li2023starcoder,qwencoder, guo2024deepseek, aixcoder} have demonstrated strong performance, thanks to extensive training on diverse datasets. To further enhance their capabilities, posting training methods like Supervised Fine-Tuning (SFT) \cite{luo2023wizardcoder, wei2023magicoder,zhang2023self} and Direct Preference Optimization \cite{qwencoder, codedpo, stepcoder, codeoptimise, plum} are commonly applied. Preference optimization approaches focus on aligning model outputs with desired outcomes by prioritizing more favorable responses over less favorable ones. 
However, existing DPO approaches fail to address one important issue: they do not directly target the most error-prone points in generated code. Errors in these high-impact parts can lead to significant quality and reliability issues in the final output. 
We aim to address this issue by focusing the preference optimization learning on these error-prone points in the generated code. 


Some fine-grained preference optimization methods \cite{rafailov2024direct, lai2024step, stepctrldpo,tdpo, cdpo} have shown strong potential in domains like mathematics, which rely heavily on natural language reasoning. Step-DPO \cite{lai2024step} and Step-Controlled DPO \cite{stepctrldpo} propose generating step-wise preference datasets to enable optimization learning based on the standard DPO loss. TDPO \cite{tdpo} enhances the DPO loss by incorporating forward KL divergence constraints at the token level, achieving fine-grained alignment for each token. cDPO \cite{cdpo} proposes a tricky method to find the critical token in the thought chain that affects overall accuracy. However, the identified tokens are typical in natural language and the method does not apply to code, which features similar overall patterns but relies on specific key elements in long reasoning processes.
However, in the context of code generation, where a small error-prone point can lead to major functional errors, these exisiting methods often struggle to construct adequate datasets or fail to achieve ideal improvements due to weak fine-grained reward signals.
To address this, we propose Focused-DPO, a framework that improves code generation by focusing on optimizing these high-impact parts. 
Our dataset construction method employs a self-generation and validation process to construct datasets that explicitly identify error-prone points, ensuring the optimization learning process directly enhances the parts of the code that matter most for overall correctness.

\section{Focused-DPO}

\begin{figure}[t]
\centering
  \includegraphics[width=\columnwidth]{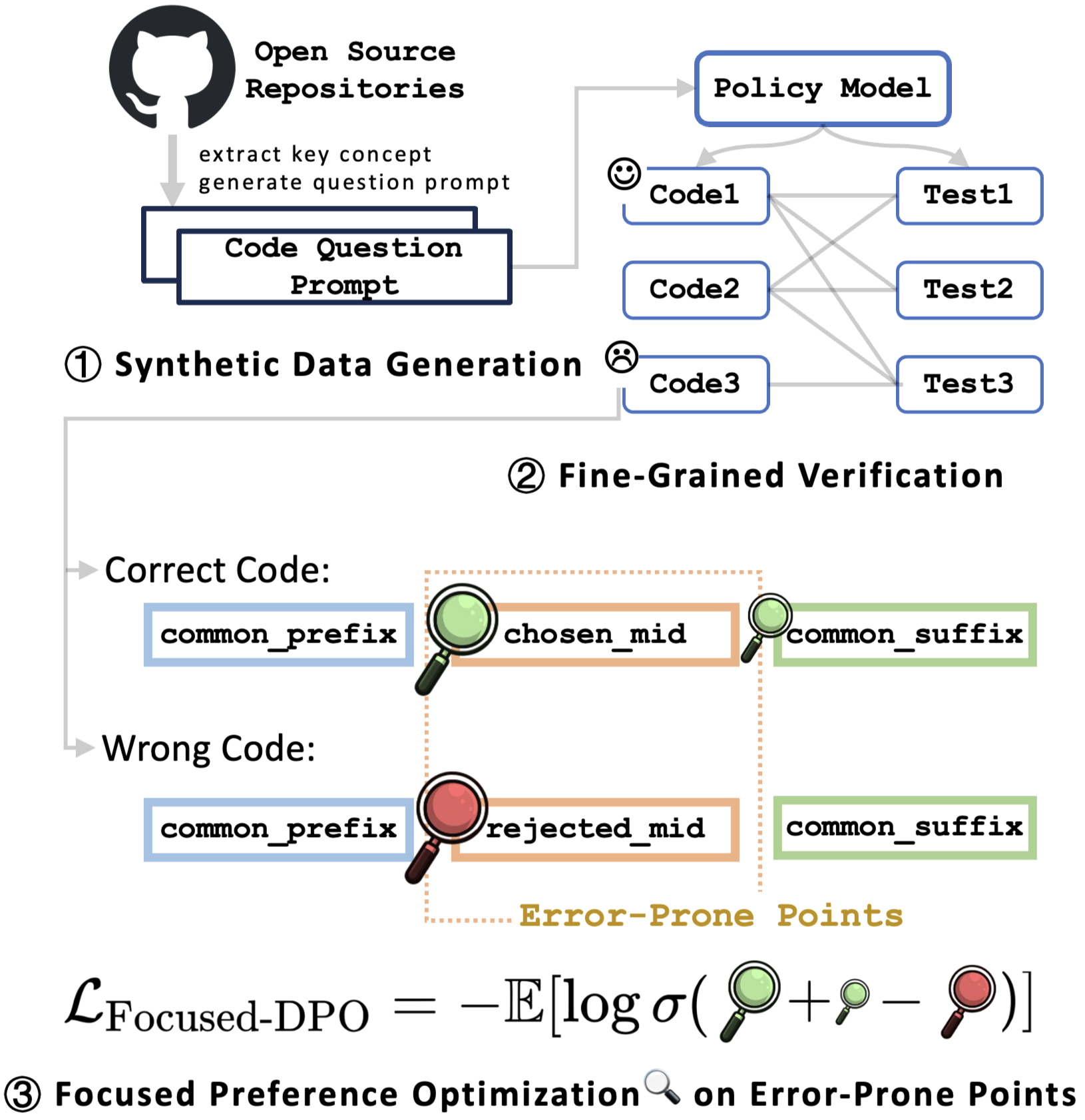}  
\caption{Overview of the Focused-DPO framework. Focused-DPO consists of three key stages: \ding{182} Generating synthetic question prompts from real-world code repositories. \ding{183} Using a policy model to simultaneously generate code and test cases, applying a page-rank algorithm to identify correct and incorrect samples and locate error-prone points using common prefixes and suffixes. \ding{184} Applying Focused-DPO, which pays special attention on error-prone points as if applying a magnifying glass for focused optimization.}
\label{fig:Method}
\end{figure}

Our proposed Focused-DPO framework aims to enhance code generation by concentrating on error-prone points through focused preference optimization. 
Building on Direct Preference Optimization, our Focused-DPO specifically targets those high-impact parts of the source code, rather than treating all code parts equally.
As illustrated in Figure \ref{fig:Method}, our method involves three main steps: 
\ding{182} \textbf{Synthetic Data Generation with Real-World Source Code} : We initiate by collecting a seed dataset from open-source code repositories and generate programming task prompts.
\ding{183} \textbf{Fine-Grained Verification to Identify Error-Prone Points} : We generate both code and tests simultaneously using a self-generation-and-validation loop. We apply a PageRank algorithm to iteratively update scores and rank the outputs, identifying correct and incorrect code samples. 
By distinguishing between similar versions of correct code and incorrect code, we locate significant parts that highly affect the final correctness and identify these parts as error-prone points, allowing for further fine-grained optimization learning.
\ding{184} \textbf{Focused Preference Optimization Learning} : We design a learning optimization algorithm specifically for these critical error-prone points. 
Using the constructed dataset, our novel training loss helps the model develop a preference for these focused parts within the code, thus optimizing performance more effectively.

\subsection{Synthetic Data Generation with Real-World Source Code}

The first step in our approach is the construction of a synthetic dataset. We collect a diverse set of programming snippets from open-source repositories to create a seed dataset. 
Similar to OSS-instruct \cite{selfoss}, we use the seed dataset to extract key programming concepts, such as algorithm design and data structure utilization. 
Then based on these concepts we generate the final prompts. 
This construction strategy allows the model to explore a broad range of scenarios. The generated question prompts are used in the following stages.

\subsection{Fine-Grained Verification to Identify Error-Prone Points}
\label{sec:dataconstruct}

To identify error-prone points, we propose a dataset construction method named \textbf{Error-Point Identification}.
Firstly, we use the policy model to simultaneously generate $k$ output codes and test cases based on the question prompts using a higher-temperature setting. In our experiment, we set $k = 10$. 
Using their execution relationships, we then adopt the ranking method from CodeDPO \cite{codedpo}, a page-rank algorithm to iteratively update scores and rank the outputs:

\begin{equation}
\resizebox{0.7\linewidth}{!}{$
\begin{split}
\text{Score}_t(c_i) &= (1 - d) \times \text{Score}_{t-1}(c_i) \\
&\quad + d \times \sum_{t_j} \text{Score}_{t-1}(t_j) \times \text{Link}(t_j, c_i)
\\
\text{Score}_t(t_j) &= (1 - d) \times \text{Score}_{t-1}(t_j) \\
&\quad + d \times \sum_{c_i} \text{Score}_{t-1}(c_i) \times \text{Link}(c_i, t_j)
\end{split}
$}
\end{equation}
Where \( d \) is the damping factor, and \( \text{Link}(t_j, c_i) \) indicates whether a code snippet \( c_i \) passes the test case \( t_j \). 
The ranking score is updated iteratively until the ranking of the code stabilizes. 

We consider the test case that the highest-ranked code correctly passes as the ground truth test case for this question. Subsequently, we split all generated codes into two categories: correct code that passes all ground truth test cases and incorrect code that does not.
For each pair consisting of a correct code sample and an incorrect code sample, we match their common prefix and suffix to decompose each code snippet into three parts: \(\texttt{common\_prefix}\), \(\texttt{mid\_chosen}\) (or \(\texttt{mid\_rej}\)), and \(\texttt{common\_suffix}\). We then define a \(\text{\textit{Diff}}\) function as follows:

\begin{equation}
\resizebox{\linewidth}{!}{$
\begin{aligned}
\text{\textit{Rank}}(\text{mid}) = & \text{Score}(\text{common\_prefix}, \text{mid}, \text{common\_suffix}), \\
\text{\textit{Diff}} =& \text{Rank}(\text{mid\_chosen}) - \text{Rank}(\text{mid\_rej}) \\
&+ \lambda * (\text{length}(\text{common\_prefix}) +  \text{length}(\text{common\_suffix})).
\end{aligned}
$}
\end{equation}

Our constructed \(\text{\textit{Diff}}\) function includes two components: \ding{182} the difference in rank between the correct and incorrect code, and \ding{183} the sum of the lengths of the common prefix and suffix, which ensures that the error-prone points are more concentrated.
We maximize \(\text{\textit{Diff}}\) to choose the \(\texttt{mid\_chosen}\) and \(\texttt{mid\_rej}\) parts that significantly impact the code's correctness, and identify these as the error-prone points. 
By focusing on error-prone points, we create training samples that directly address the parts of the code that have significantly impact on correctness. 
For each policy model, we apply necessary filtering to the generated data, resulting in a final dataset containing 5,000 training samples and 1,000 validation samples. Table \ref{tab:training-dataset-statistics} presents an example of data statistics.

\subsection{Focused Preference Optimization Learning}
\label{sec:methodloss}

The core of our method lies in modifying the Direct Preference Optimization (DPO) framework to better enhance code generation by focusing on error-prone points of the code. 
Given a pairwise preference dataset \(\mathcal{D} = \{(x_i, y^{chosen}_i, y^{rej}_i)\}_{i=1}^M\), the standard DPO loss \cite{rafailov2024direct} is expressed as:
\begin{equation}
\resizebox{\linewidth}{!}{$
\ell_{\text{DPO}} = -\mathbb{E}_{(x, y^{chosen}, y^{rej}) \sim \mathcal{D}} \left[ \log \sigma \left(\phi(x, y^{chosen}) - \phi(x, y^{rej}) \right)\right],
$}
\end{equation}
where \(\phi(x, y)\) is an implicit reward function. The reward function is defined as:
\begin{equation}
\resizebox{\linewidth}{!}{$
\phi(x, y) = \beta \cdot \log \frac{\pi_\theta(y|x)}{\pi_{\text{ref}}(y|x)} + \underbrace{\beta \cdot \log Z(x)}_{\text{this term can ultimately be reduced}}
$}
\end{equation}
where \(\pi_\theta(y|x)\) represents the probability of a generated response \(y\) under the policy model, and \(\pi_{\text{ref}}(y|x)\) is the probability under a reference model, typically the SFT baseline. The goal of DPO loss is to maximize reward difference between the preferred and non-preferred samples.

\paragraph{Reward Function Modification}
In its original form, the DPO reward \(\phi(x, y)\) is calculated over the entirety of the sample \(y\), treating all parts of the code equally. 
However, in the context of code generation, not all parts of the code contribute equally to correctness. 
Building on our observation that the middle part (\(\texttt{mid}\)) of code—the error-prone point we identify in Section \ref{sec:dataconstruct}—should receive more attention, we restructure the reward to reflect the relative importance of different code parts.
The reward function is modified to weight the \(\texttt{mid}\) part more heavily, reflecting its critical contribution to the correctness of the code. For the preferred sample, the reward function becomes:

\begin{equation}
\resizebox{0.7\linewidth}{!}{$
\begin{split}
&\phi_{\text{chosen}}(x, y) = \beta  \cdot \Big( \log \frac{\pi_\theta(\texttt{prefix}|x)}{\pi_{\text{ref}}(\texttt{prefix}|x)} \\
& + w_{\text{focused}} \cdot \log \frac{\pi_\theta(\texttt{mid}|x, \texttt{prefix})}{\pi_{\text{ref}}(\texttt{mid}|x, \texttt{prefix})} \\
& + \log \frac{\pi_\theta(\texttt{suffix}|x, \texttt{prefix}, \texttt{mid})}{\pi_{\text{ref}}(\texttt{suffix}|x, \texttt{prefix}, \texttt{mid})} \Big)
\end{split}
$}
\end{equation}

Where \(w_{\text{focused}}\) is a weight that amplifies the importance of the \(\texttt{mid}\) part.

For the non-preferred sample, we adopt a similar structure but introduce an adjustment to further downweight the contribution of the \(\texttt{suffix}\). This adjustment is based on our observation that regardless of whether the \(\texttt{mid}\) part contains errors, the content of the \(\texttt{suffix}\) is often the same or similar. 
Our results in Section \ref{sec:experimentrq1} show that the correlation between the \(\texttt{suffix}\) and the overall accuracy of the final code is low, making it less significant in the reward calculation. The reward becomes:
\begin{equation}
\resizebox{0.7\linewidth}{!}{$
\begin{split}
&\phi_{\text{rej}}(x, y) = \gamma  \cdot \Big( \log \frac{\pi_\theta(\texttt{prefix}|x)}{\pi_{\text{ref}}(\texttt{prefix}|x)} \\
& + w_{\text{focused}} \cdot \log \frac{\pi_\theta(\texttt{mid}|x, \texttt{prefix})}{\pi_{\text{ref}}(\texttt{mid}|x, \texttt{prefix})} \Big)
\end{split}
$}
\end{equation}

\paragraph{Final Loss Function}


Substituting the modified rewards for the preferred (\(y^{chosen}\)) and non-preferred (\(y^{rej}\)) examples into the original DPO loss and simplifying by canceling common terms, we can obtain that:
\begin{equation}
\resizebox{\linewidth}{!}{$ 
\begin{aligned} 
\Delta_{\texttt{reward}} 
= & \phi_{\text{chosen}}(x, y^{chosen}) - \phi_{\text{rej}}(x, y^{rej}) \\
= & 
\underbrace{
\begin{aligned}
\sum_{j=k+1}^{m} \beta \cdot w_{\text{focused}} \cdot \log \frac{\pi_\theta(t_j^{\text{(mid\_chosen)}} | x, t_{0:k}^\text{(prefix)}, t_{k+1:j-1}^\text{(mid\_chosen)})}{\pi_{\text{SFT}}(t_j^{\text{(mid\_chosen)}} | x, t_{0:k}^\text{(prefix)}, t_{k+1:j-1}^\text{(mid\_chosen)})} \\
- \sum_{j=k+1}^{n} \beta \cdot w_{\text{focused}} \cdot \log \frac{\pi_\theta(t_j^{\text{(mid\_rej)}} | x, t_{0:k}^\text{(prefix)}, t_{k+1:j-1}^\text{(mid\_rej)})}{\pi_{\text{SFT}}(t_j^{\text{(mid\_rej)}} | x, t_{0:k}^\text{(prefix)}, t_{k+1:j-1}^\text{(mid\_rej)})}
\end{aligned}
}_{\Delta_{\texttt{mid}}} \\
& +
\underbrace{
\sum_{j=m+1}^{L_1} \beta \cdot \log \frac{\pi_\theta(t_j^{\text{(suffix)}} | x, t_{0:k}^\text{(prefix)}, t_{k+1:m}^\text{(mid\_chosen)}, t_{m+1:j-1}^\text{(suffix)})}{\pi_{\text{SFT}}(t_j^{\text{(suffix)}} | x, t_{0:k}^\text{(prefix)}, t_{k+1:m}^\text{(mid\_chosen)}, t_{m+1:j-1}^\text{(suffix)})}
}_{\Delta_{\texttt{suffix}}} \\
= & \Delta_{\texttt{mid}} + \Delta_{\texttt{suffix}}
\end{aligned}
$}
\end{equation}

So the final loss function for Focused-DPO is expressed as:
\begin{equation}
\resizebox{0.9\linewidth}{!}{$
\begin{split}
\mathcal{L}&_{\text{Focused-DPO}}(\pi_\theta; \pi_\text{SFT}) = \\
& -\mathbb{E}_{(x, y^{chosen}, y^{rej}) \sim \mathcal{D}} \left[
\log \sigma \left( \Delta_{\texttt{mid}} + \Delta_{\texttt{suffix}} \right)
\right],
\end{split}
$}
\end{equation}

The terms \(\Delta_{\texttt{mid}}\) and \(\Delta_{\texttt{suffix}}\) capture the weighted differences in the probabilities of critical parts between the preferred and non-preferred samples, with greater emphasis focused on the \(\texttt{mid}\) parts, which is the error-prone point.

Through this modification, Focused-DPO shifts the focus of optimization toward the error-prone point in the code. 
By increasing the weight of these parts in the reward calculation, our framework ensures that the model prioritizes improvements where they matter most, leading to higher-quality and more reliable code generation.

\section{Experiment Setup}
\label{sec:experimentsetup}
We aim to answer the following research questions:

\paragraph{RQ1: Are there error-prone points in generated code that significantly affect the correctness of the output?}
This question addresses the core motivation behind Focused-DPO. To investigate this, we construct the validation dataset following Section \ref{sec:dataconstruct}.
This setup provides empirical evidence supporting the theoretical underpinnings of our Focused-DPO.

\paragraph{RQ2: Can Focused-DPO improve the generation quality of code models, even those that have already been heavily post-trained with alignment techniques such as standard DPO?}
To explore this, we evaluate Focused-DPO on several widely-used code generation benchmarks, including HumanEval \citep{chen2021evaluating}, HumanEval+ \citep{liu2024your}, MBPP \citep{austin2021program}, MBPP+, and LiveCodeBench \cite{jain2024livecodebench}. 




\paragraph{RQ3: How do different components of the Focused-DPO loss formulation affect model performance?}
Ablation studies include evaluating our dataset construction method, as well as key components in our loss formulation.


\subsection{Baselines}
\label{sec:setupLLM}

We evaluate several widely used large language models (LLMs) in the code generation domain.
For \textbf{\textit{base models}}, we apply Focused-DPO to \textbf{DeepSeekCoder-base-6.7B)} \citep{guo2024deepseek} and \textbf{Qwen2.5-Coder-7B} \citep{qwencoder}. 
For \textbf{\textit{instruct models}} , we evaluate on \textbf{Magicoder-S-DS-6.7B} \citep{wei2023magicoder} and \textbf{DeepSeekCoder-instruct-6.7B}, which are post-trained from \textit{DeepSeekCoder-base-6.7B} with large-scale SFT. We further evaluate \textbf{Qwen-2.5-Coder-Instruct-7B}, which is post-trained from \textit{Qwen2.5-Coder-7B} on million-level datasets with SFT and DPO.

We compare against several widely used training techniques, including: 
\textbf{SFT}, \textbf{standard DPO}, \textbf{Step-DPO} \cite{lai2024step}, \textbf{TDPO} \cite{tdpo}.  
SFT trains models only with positive samples, while the other methods utilize a pairwise dataset of preferred and rejected samples.

\subsection{Training and Inference Settings}

For each backbone LLM, we sample 10 code candidates and corresponding test cases for each problem prompt using \texttt{temperature=1.5}. An example of data statistics is in Table \ref{tab:training-dataset-statistics}. 
Our analysis shows this configuration results in a stable ranking score and ensures diversity. We focus on Python-based datasets given its widespread use.
For training, we train for 10 epochs on 8 NVIDIA V100 GPUs and select the best-performing checkpoint based on the lowest validation loss. We set $w_{focused} = 2$ in our experiments.
We use a learning rate of \(5 \times 10^{-6}\) with a linear scheduler and warm-up. 
We employ greedy search during inference. 

\section{Results and Analyses}

\subsection{Exploration of Error-Prone Points in Code (RQ1)}
\label{sec:experimentrq1}

We conduct experiments to validate our motivation:

\noindent \ding{182} Correlation analysis confirms that \textbf{error-prone points in the code significantly impact correctness}, whereas other code parts have minimal effect.

\noindent \ding{183} Generation experiments show that \textbf{continuing at these points with different content leads to significant differences in overall correctness}.

\noindent \ding{184} Observations reveal that \textbf{existing code models perform suboptimally at these points}.
\paragraph{Correlation Between Different Code Parts and Final Correctness}

Utilizing the dataset construction pipeline described in Section \ref{sec:dataconstruct}, we evaluate the validation dataset based on \textit{Qwen2.5-Coder-Instruct-7B}. 
We analyze the relationship between \(\texttt{prefix}\), \(\texttt{suffix}\), two types of \(\texttt{mid}\) parts, and the final code correctness, as presented in Table \ref{tab:segment-frequencies-and-phi}.

\begin{table}[h]
    \centering
    \resizebox{\linewidth}{!}{
    \begin{tabular}{lcc|c}
        \toprule
        \textbf{Segment} & \textbf{Correct} & \textbf{Incorrect} & \textbf{Phi Coefficient}  \\
        \midrule
        Common Prefix                    & 0.7907 & 0.7325 & \cellcolor{yellow!30}0.0683 \\
        Common Suffix                    & 0.8479 & 0.7864 & \cellcolor{yellow!30}0.0796  \\ \midrule
        Common Prefix + Chosen Mid       & 0.6367 & 0.0911 & \cellcolor{green!30}0.5651  \\
        Common Prefix + Reject Mid       & 0.0116 & 0.5575 & \cellcolor{red!30}-0.6085  \\
        \bottomrule
    \end{tabular}
    }
    \caption{Relationships between the \(\texttt{prefix}\), \(\texttt{suffix}\), and the two types of \(\texttt{mid}\) parts with the final code correctness. The table includes the frequency of each part in correct and incorrect code, as well as their correlation coefficients with overall code correctness.}
    \label{tab:segment-frequencies-and-phi}
\end{table}

Results in Table \ref{tab:segment-frequencies-and-phi} show that \texttt{common\_prefix + chosen\_mid} appears much more frequently in correct solutions, while \texttt{common\_prefix + rej\_mid} is prevalent in incorrect solutions. 
This confirms the critical influence of the \texttt{mid} part, with strong positive and negative correlations respectively, affirming the existence of error-prone points in generated code.
In contrast, we find that the prefix and suffix parts have little relation to the correctness of the final answer. It is important to note that in incorrect code, despite the errors in the \texttt{mid} section, the following suffix is not a significant cause of the errors. 
This observation justifies our decision to exclude the suffix in the reward modification in Section \ref{sec:methodloss}.
These findings provide empirical evidence supporting our hypothesis that focusing on these error-prone points is essential to enhance model performance, which is the core motivation behind our Focused-DPO framework.

\paragraph{Accuracy of Continuation at Error-Prone Points}

We further generate 20 code solutions based on different contents at error-prone points, to explore the correctness of the final code generated under different conditions in Table \ref{tab:pass-rates-mid}. 

\begin{table}[h]
    \centering
    \resizebox{\linewidth}{!}{
    \begin{tabular}{lcccc}
        \toprule
        \textbf{Based on Input} & \textbf{pass@1} & \textbf{pass@3} & \textbf{pass@5} & \textbf{pass@10} \\
        \midrule
        Common Prefix + Chosen Mid   & 0.9002 & 0.9532 & 0.9688 & 0.9871 \\
        Common Prefix + Reject Mid & 0.0317 & 0.0633 & 0.0810 & 0.1159 \\
        \bottomrule
    \end{tabular}
    }
    \caption{Pass rates based on different content at error-prone points.}
    \label{tab:pass-rates-mid}
\end{table}

The pass rates shown in Table \ref{tab:pass-rates-mid} highlight a striking contrast: using \texttt{chosen\_mid} at error-prone points results in significantly higher pass rates, reaching around 90\% at pass@1, compared to just over 3\% for the \texttt{rej\_mid} version. This demonstrates the critical importance of accurate content in the error-prone points for determining the correctness of the final generated code.

Based on the above results, we have noticed that \textbf{the generated content at the error-prone points significantly affects the final outcomes}. 
This leads to a question: \textit{how do current code generation models behave at these error-prone points?}

\paragraph{Generation Preferences at Error-Prone Points in Code Models}

\begin{figure}[h]
\centering
  \includegraphics[width=\columnwidth]{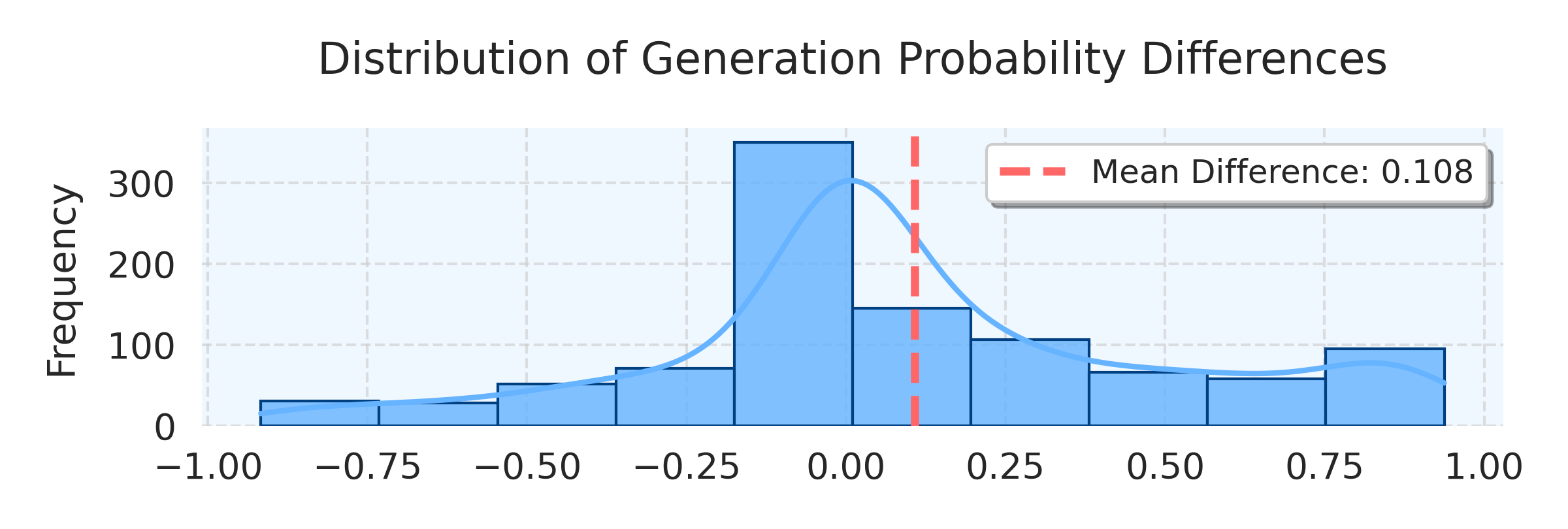}  
\caption{Generation probability difference \((p(\text{chosen\_mid}) - p(\text{rej\_mid}))\) with input.}
\label{fig:motivationprob}
\end{figure}

We further analyze the \textit{Qwen-2.5-Coder-Instruct-7B}, which has been post-trained on million-level datasets using SFT and DPO. We examine the generation preferences of this heavily post-trained model at error-prone points. Specifically, we calculate the probability difference between generating \texttt{chosen\_mid} and \texttt{rej\_mid} when given the \texttt{common\_prefix} as input. The distribution of the difference is shown in Figure \ref{fig:motivationprob}. The model exhibits little to no clear preference, indicating that existing code generation models lack effective generation capability at these error-prone points.
Through this exploration, we confirm that focused preference optimization of error-prone points is crucial for improving the accuracy of code models, addressing RQ1.

\subsection{Main Results (RQ2)}

\paragraph{Results on benchmarks}
Tables \ref{tab:main-results-benchmark1} and \ref{tab:livecodebenchresults} summarize the performance of Focused-DPO compared to various baselines, including standard DPO, Step-DPO, TDPO, and SFT. 
Note that the formulas for standard DPO and Step-DPO are identical, making them equivalent.
The relative improvements (\textit{Rel}) are reported for a clearer comparison.

\begin{table}[h]
    \centering

    \resizebox{\linewidth}{!}{
    
    \begin{tabular}{lcccc}
        \toprule
        \textbf{Model}                & \textbf{HumanEval} & \textbf{HumanEval+} & \textbf{MBPP} & \textbf{MBPP+} \\
        \bottomrule
        \textit{Instruct Model} & & & & \\
        \toprule
        \textbf{Qwen2.5-coder-instruct-7B}        & 0.915              & 0.841              & 0.828         & 0.714          \\
        + Our Focused-DPO                   & \textbf{0.927}     & \textbf{0.878}     & \textbf{0.847} & \textbf{0.762} \\
        \textit{Relative Improvement}    & 1.29\%             & 4.41\%             & 2.24\%        & 6.71\%         \\
        \midrule
        DPO / Step-DPO                & 0.921              & 0.854              & 0.841         & 0.743          \\
        Token-DPO                          & 0.927              & 0.872              & 0.833         & 0.751          \\
        SFT                           & 0.927              & 0.872              & 0.833         & 0.717          \\
        \midrule
                \textbf{DeepSeekCoder-instruct-6.7B}        & 0.774              & 0.701              & 0.751         & 0.659          \\
        + Our Focused-DPO                   & \textbf{0.823}     & \textbf{0.732}     & \textbf{0.765} & \textbf{0.669} \\
        \textit{Relative Improvement}    & 6.35\%             & 4.38\%             & 1.80\%        & 1.56\%         \\
        \midrule
        DPO / Step-DPO                & 0.787              & 0.713              & 0.751         & 0.661          \\
        Token-DPO                          & 0.799              & 0.726              & 0.751         & 0.661          \\
        SFT                           & 0.787              & 0.726              & 0.759         & 0.667          \\
        \midrule
                \textbf{MagiCoder-S-DS-6.7B}        & 0.732              & 0.683              & 0.767         & 0.667          \\
        + Our Focused-DPO                   & \textbf{0.823}     & \textbf{0.744}     & \textbf{0.794} & \textbf{0.698} \\
        \textit{Relative Improvement}    & 12.50\%            & 8.93\%             & 3.45\%        & 4.76\%         \\
        \midrule
        DPO / Step-DPO                & 0.762              & 0.701              & 0.772         & 0.675          \\
        Token-DPO                          & 0.811              & 0.732              & 0.780         & 0.680          \\
        SFT                           & 0.738              & 0.701              & 0.762         & 0.653          \\
        \bottomrule
        \textit{Base Model} & & & & \\
        \toprule
\textbf{Qwen2.5-coder-base}        & 0.835              & 0.787              & 0.794         & 0.683          \\
        + Our Focused-DPO                   & \textbf{0.884}     & \textbf{0.829}     & \textbf{0.817} & \textbf{0.704} \\
        \textit{Relative Improvement}    & 5.89\%             & 5.37\%             & 2.95\%        & 3.03\%         \\
        \midrule
        DPO / Step-DPO                & 0.848              & 0.799              & 0.802         & 0.688          \\
        Token-DPO                          & 0.866              & 0.799              & 0.815         & 0.690          \\
        SFT                           & 0.848              & 0.805              & 0.802         & 0.688          \\
        \midrule
                \textbf{DeepSeekCoder-base-6.7B}        & 0.476              & 0.396              & 0.702         & 0.566          \\
        + Our Focused-DPO                   & \textbf{0.518}     & \textbf{0.427}     & \textbf{0.717} & \textbf{0.574} \\
        \textit{Relative Improvement}    & 8.89\%             & 7.79\%             & 2.13\%        & 1.43\%         \\
        \midrule
        DPO / Step-DPO                & 0.488              & 0.396              & 0.709         & 0.569          \\
        Token-DPO                          & 0.500              & 0.421              & 0.717         & 0.574          \\
        SFT                           & 0.488              & 0.396              & 0.704         & 0.566          \\
        \bottomrule
    \end{tabular}
    }
        \caption{Pass Rate on HumanEval(+), MBPP(+)}
    \label{tab:main-results-benchmark1}
\end{table}

\begin{table}[h]
    \centering
    \resizebox{\linewidth}{!}{
    
    \begin{tabular}{lcccc}
        \toprule
        \textbf{Model}                & \textbf{Easy} & \textbf{Medium} & \textbf{Hard} & \textbf{Average} \\
        \bottomrule
        \textit{Instruct Model} & & & & \\
        \toprule
        \textbf{Qwen2.5-coder-instruct-7B} & 0.692       & 0.220         & 0.034         & 0.312        \\
        + Our Focused-DPO                   & \textbf{0.735} & \textbf{0.242} & \textbf{0.048} & \textbf{0.339} \\
        \textit{Relative Improvement}       & 6.22\%         & 10.04\%       & 42.86\%       & 8.44\%        \\
        \midrule
        DPO / Step-DPO                & 0.685       & 0.233         & 0.019         & 0.310        \\
        Token-DPO                          & 0.706       & 0.239         & 0.037         & 0.325        \\
        SFT                           & 0.670       & 0.208         & 0.015         & 0.295        \\
        \midrule
        \textbf{DeepSeekCoder-instruct-6.7B} & 0.453       & 0.091         & 0.009         & 0.181        \\
        + Our Focused-DPO                   & \textbf{0.477} & \textbf{0.106} & \textbf{0.019} & \textbf{0.197} \\
        \textit{Relative Improvement}       & 5.30\%         & 15.89\%       & 108.33\%      & 8.87\%        \\
        \midrule
        DPO / Step-DPO                & 0.462       & 0.094         & 0.007         & 0.184        \\
        Token-DPO                          & 0.470       & 0.100         & 0.019         & 0.192        \\
        SFT                           & 0.462       & 0.094         & 0.004         & 0.183        \\
        \midrule
        \textbf{MagiCoder-S-DS-6.7B} & 0.481       & 0.107         & 0.001         & 0.193        \\
        + Our Focused-DPO                   & \textbf{0.513} & \textbf{0.118} & \textbf{0.019} & \textbf{0.213} \\
        \textit{Relative Improvement}       & 6.56\%         & 10.12\%       & 1751.85\%     & 10.10\%       \\
        \midrule
        DPO / Step-DPO                & 0.491       & 0.109         & 0.004         & 0.198        \\
        Token-DPO                          & 0.505       & 0.118         & 0.015         & 0.209        \\
        SFT                           & 0.498       & 0.112         & 0.004         & 0.201        \\
        \bottomrule
        \textit{Base Model} & & & & \\
        \toprule
        \textbf{Qwen2.5-coder-base-7B}       & 0.567       & 0.150         & 0.017         & 0.241        \\
        + Our Focused-DPO                   & \textbf{0.595} & \textbf{0.175} & \textbf{0.030} & \textbf{0.264} \\
        \textit{Relative Improvement}       & 5.00\%         & 16.47\%       & 77.78\%       & 9.23\%        \\
        \midrule
        DPO / Step-DPO                & 0.577       & 0.151         & 0.015         & 0.244        \\
        Token-DPO                          & 0.584       & 0.163         & 0.022         & 0.253        \\
        SFT                           & 0.584       & 0.157         & 0.022         & 0.251        \\
        \midrule
        \textbf{DeepSeekCoder-base-6.7B} & 0.399       & 0.074         & 0.004         & 0.155        \\
        + Our Focused-DPO                   & \textbf{0.423} & \textbf{0.085} & \textbf{0.011} & \textbf{0.169} \\
        \textit{Relative Improvement}       & 6.00\%         & 14.31\%       & 177.78\%      & 9.24\%        \\
        \midrule
        DPO / Step-DPO                & 0.412       & 0.079         & 0.004         & 0.161        \\
        Token-DPO                          & 0.419       & 0.079         & 0.004         & 0.164        \\
        SFT                           & 0.419       & 0.082         & 0.007         & 0.166        \\
        \bottomrule
    \end{tabular}
    }
    \caption{Pass Rate on LiveCodeBench}
    \label{tab:livecodebenchresults}
\end{table}

As shown in Table \ref{tab:main-results-benchmark1}, Focused-DPO consistently outperforms the baseline models across all benchmarks. 
On the HumanEval(+) and MBPP(+) benchmarks, Focused-DPO improves relative accuracy by 4.79\% on average over the baseline.
We also evaluate on LiveCodeBench, a challenging benchmark that features iteratively updated, competition-level programming problems sourced from platforms such as LeetCode. The benchmark is divided into three levels of difficulty: Easy, Medium, and Hard. 
Focused-DPO achieves consistent improvements across all difficulty levels of LiveCodeBench. Notably, on the hardest category (\textit{Hard}), Focused-DPO can achieve huge relative performance.
Focused-DPO entirely outperforms other advanced preference optimization baselines such as Step-DPO and TDPO. These findings highlight the effectiveness of Focused-DPO in challenging code generation scenarios, where optimization on error-prone points of code plays a crucial role in determining final correctness.

\paragraph{Enhancing Heavily Post-trained Models}

Focused-DPO can significantly enhance the performance of code models that have already undergone extensive post-training. As demonstrated in Table \ref{tab:dsSFTDPO}, models like \textit{Qwen2.5-Coder-instruct}, which have been meticulously optimized using millions of data points from SFT and DPO processes, still exhibit substantial improvements with our Focused-DPO framework.
To further illustrate Focused-DPO's benefits on heavily post-trained models, we conducted an extensive initial DPO training phase. 
Following the methodology from CodeDPO, we used the model \textit{DeepSeekCoder-base-6.7} and a large-scale dataset with 93k samples for DPO training, continued until full convergence.
We then apply Focused-DPO for further experiments. This allows us to explore the extent to which Focused-DPO could drive additional improvements, even in models already trained by intensive post-training processes. 

\begin{table}[h]
\centering
\resizebox{\linewidth}{!}{
\begin{tabular}{l|c|c|c|c}
\toprule
\textbf{Model} & \textbf{HumanEval} & \textbf{HumanEval+} & \textbf{MBPP} & \textbf{MBPP+} \\
\midrule
DeepSeekCoder-base-6.7B & 0.4760 & 0.3960 & 0.7020 & 0.5660 \\
\midrule
+ SFT Stage & 0.7317 & 0.6829 & 0.7672 & 0.6667 \\
 \textit{(with MagiCoder-OSS-instruct)} & & & & \\
\midrule
+ First DPO Stage & 0.8354 & 0.7622 & 0.8070 & 0.7093 \\
 \textit{(with CodeDPO-OSS-instruct)} & & & & \\
\midrule
+ Focused-DPO & \textbf{0.8719} & \textbf{0.7926} & \textbf{0.8227} & \textbf{0.7275} \\
\bottomrule
\end{tabular}
}
\caption{Performance of DeepSeekCoder-6.7B at different training stages. The stages include base model, SFT with MagiCoder, first DPO with CodeDPO, and our Focused-DPO. Focused-DPO achieves additional improvements even after high-quality post-training.}
\label{tab:dsSFTDPO}
\end{table}

As shown in Table \ref{tab:dsSFTDPO}, we start from the base model and progressively incorporate the SFT stage \cite{wei2023magicoder} and the first DPO stage \cite{codedpo}. 
Finally, applying our Focused-DPO leads to the highest pass rates achieved. 
These results demonstrate that Focused-DPO effectively boosts the performance of models that have already been extensively post-trained and optimized through previous stages. 
We further evaluate how Focused-DPO enhances the quality of error-prone points in Appendix \ref{sec:improveerrorpronepoints}.

\subsection{Ablation Study (RQ3)}


\paragraph{Dataset Construction Ablations}

Focused-DPO includes an automated data construction and Error-Prone Identification process. 
We perform ablation experiments on the dataset construction methods in Table \ref{tab:dataset-construction-ablation}.
We design two alternative approaches:
\ding{182} The \textit{Step-DPO strategy} \cite{lai2024step} constructs datasets by considering only the common prefix parts, with the rest treated as Error-Prone Points for training.
\ding{183} Using a \textit{git-diff tool} \footnote{\url{https://git-scm.com/docs/git-diff}}, we construct datasets where the differences covered by the diff were treated as Error-Prone Points, with the parts following the final diff difference treated as the suffix.
Note that Step-DPO dataset construction method is closely tied to the formulation of the Step-DPO loss function, leading to consistent outcomes between the two. However, we observe that Step-DPO performs suboptimally on code generation tasks. 
In contrast, the current dataset construction method used in Focused-DPO, which employs a simple yet effective Error-Prone Identification strategy, achieves the best experimental results.

\paragraph{Loss Function Ablations}
Our Focused-DPO has made appropriate modifications to the calculation of positive and negative rewards. 
We carry out ablation experiments in Table \ref{tab:dataset-construction-ablation}, including trying different values of $w_{focused}$ and various treatments of the suffix in the reward function.
Our findings indicate that increasing or decreasing $w_{focused}$ leads to a decline in model performance, suggesting that the current value of $w_{focused}$ is optimal. 
Additionally, we observe that including the suffix part in the reward function results in degraded performance. Through detailed analysis in Section \ref{sec:experimentrq1}, the suffix in incorrect code does not exhibit strong correlations with the overall accuracy. 
These experiments validate the practical advantages of the design choices in our loss function.

\begin{table}[h!]
    \centering
    \resizebox{\linewidth}{!}{
    \begin{tabular}{lcc}
        \toprule
        \textbf{Dataset Construction} & \textbf{HumanEval / HumanEval+} & \textbf{MBPP / MBPP+} \\
        \midrule
        Focused-DPO    & & \\
        \makecell[l]{Error Prone Identification}     & \textbf{0.9268 / 0.8780} & \textbf{0.8466 / 0.7619} \\
        \midrule
        Step-DPO Strategy                      & 0.9207 / 0.8537          & 0.8413 / 0.7434          \\
        Diff-based Strategy                    & 0.9268 / 0.8598          & 0.8439 / 0.7539          \\
        \bottomrule
        \toprule
        \textbf{Loss Function Setting} & \textbf{HumanEval / HumanEval+} & \textbf{MBPP / MBPP+} \\
        \midrule
        Focused-DPO    & & \\
        \makecell[l]{$w_{focused} = 2$,\\ No Suffix in Reject Reward}     & \textbf{0.9268 / 0.8780} & \textbf{0.8466 / 0.7619} \\
        \midrule
        \textit{Decrease Weight}    & & \\
        $w_{focused} = 1$         & 0.9268 / 0.8720          & 0.8386 / 0.7487          \\
        \midrule
        \textit{Increase Weight}    & & \\
        $w_{focused} = 3$             & 0.9268 / 0.8720          & 0.8439 / 0.7566          \\
        $w_{focused} = 5$              & 0.8963 / 0.7683          & 0.8201 / 0.6878          \\
        \midrule
        Suffix in Reject Reward & 0.9268 / 0.8659          & 0.8413 / 0.7487          \\
        \bottomrule
    \end{tabular}
    }
    \caption{Dataset Construction and Loss Function Ablation Results based on \textit{Qwen2.5-Coder-Instruct-7B}}
    \label{tab:dataset-construction-ablation}
    \vspace{-15pt}
\end{table}



\section{Discussions}
\label{sec:discussion}

\subsection{Error-Prone Points Identification}

In our Focused-DPO method, we introduce a dataset construction technique called \textbf{Error Prone Identification} to automatically identify error-prone points in generated code. To assess the correctness of the code, we employ a self-generation-and-validation mechanism based on PageRank, which captures the relative quality of different code snippets \cite{codedpo}. We are not like approaches such as Magicoder \cite{wei2023magicoder}, which directly use all test cases as ground truth. In our experiments we use the policy model to generate datasets. Since the policy model's generation quality is not as robust as that of more powerful models like GPT-4 (used in Magicoder), the PageRank-based method allows us to automatically filter out lower-quality test cases (those with lower scores after iteration), thereby ensuring higher overall dataset quality.

We find that different models exhibit varying levels of accuracy across different problems. Therefore, for each model's training dataset, we performed necessary filtering by removing code problems with excessively high or low accuracy rates, ensuring a consistent number of code problems in the final dataset. Moreover, we observe that models tend to exhibit similarities in error-prone points when solving the same problems. For example, when comparing the error-prone points identified by \textit{DeepSeekCoder-instruct-6.7B} and \textit{Qwen2.5-Coder-instruct-7B} models on the same set of programming problems, we found a 32\% overlap. This indicates that there are commonalities in the error-prone points across different models.

In our ablation studies, we compare error-prone points constructed using the \textit{git-diff} method and the \textit{Step-DPO} method, noting slight differences in the final results. 
Balancing effectiveness and efficiency, we use the method based on \textit{prefix} and \textit{suffix}, which allows us to identify error-prone points in generated code in a simple yet effective manner. We plan to further explore more identification strategies in future work.

\subsection{Improvement in Error-Prone Points}
\label{sec:improveerrorpronepoints}
\begin{figure}[h]
\centering
  \includegraphics[width=\columnwidth]{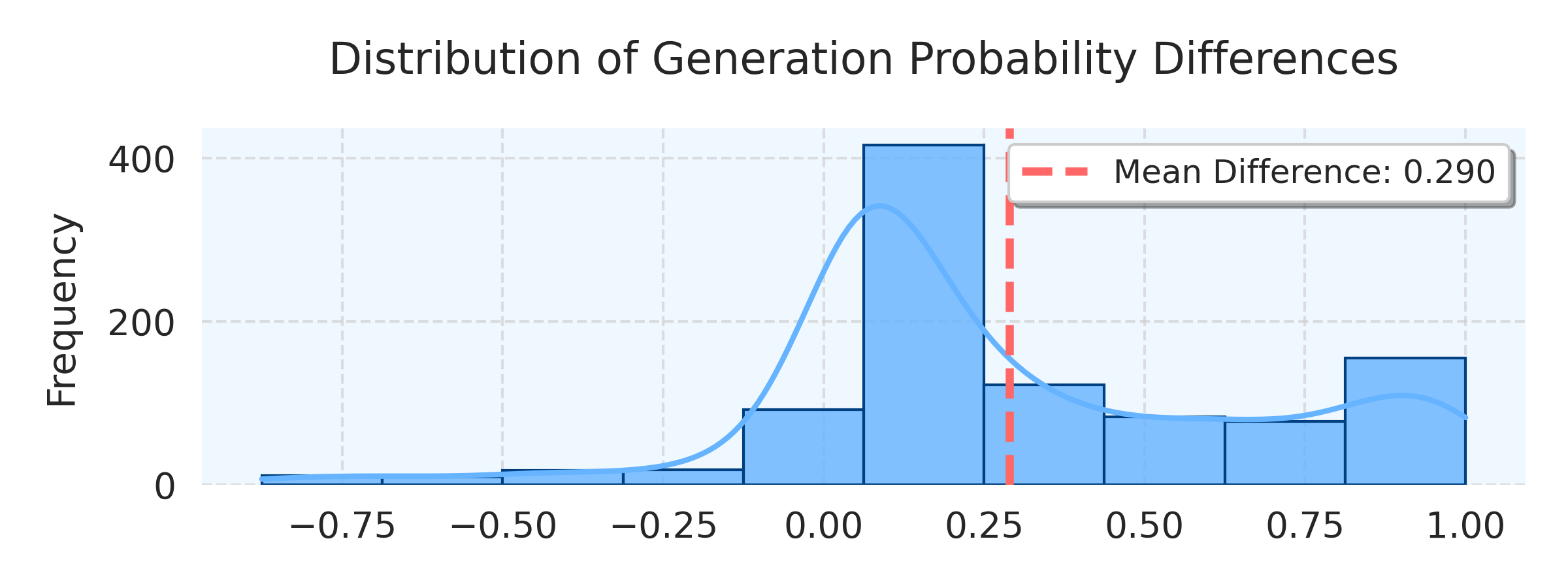}  
\caption{Generation Probability Difference (p(chose\_mid) - p(reject\_mid)) after Focused-DPO.}
\label{fig:diffaftertraining}
\vspace{-15pt}
\end{figure}

We further evaluate how Focused-DPO enhances the quality in error-prone points. 
Using our validation dataset (Table \ref{tab:dataset-statistics}), we measure the model's performance on these error-prone parts. 
The generation probability difference between \textit{chosen\_mid} and \textit{reject\_mid} in error-prone points is illustrated in Figure \ref{fig:diffaftertraining} for \textit{Qwen2.5-Coder-Instruct-7B} model.

Compared to pre-Focused-DPO results (Figure \ref{fig:motivationprob}), Focused-DPO demonstrates a strong preference for generating more accurate code at error-prone points. 
This improvement is particularly critical in complex coding tasks, where precise decisions in error-prone points directly impact the correctness of the generated code. 
For instance, on the \textbf{LiveCodeBench-Hard} dataset—which consists of challenging, dynamically problems—Focused-DPO achieves a significant improvement of 42.8\% in correctness for the \textit{Qwen2.5-Coder-Instruct} model. 
Notably, on this dataset, Focused-DPO achieves performance on par with \textbf{GPT-4o}, highlighting its ability to address difficult code generation tasks effectively.

\section{Conclusion}


We propose Focused-DPO, a framework that improves code generation by focusing on error-prone points. 
These critical parts significantly impact overall program correctness. 
Focused-DPO improves Direct Preference Optimization by prioritizing these points, using our Error-Point Identification method to create datasets without costly human annotations.
Evaluations show Focused-DPO reduces errors and improves quality, even in heavily post-trained models.  
This research highlights benefits of focusing on fine-grained preference optimization in AI-driven software development.

\section{Acknowledgement}

This research is supported by the National Natural Science Foundation of China under Grant
Nos. 62192731,  62192730, 62192733, 62072007, the National Key R\&D Program under Grant No. 2023YFB4503801, and the Major Program (JD) of Hubei Province (No.2023BAA024).
\footnote{This work is based on our prior findings in CodeDPO \cite{codedpo}. We are deeply grateful to Jingjing Xu, Jing Su, and Jun Zhang for their valuable discussions.}

\section*{Limitation}

Despite the contributions of our work, there are several limitations that we aim to address in future research:

\paragraph{Comparison with Advanced RL Techniques}
While our study demonstrates the effectiveness of Focused-DPO, we do not extensively compare it with other advanced reinforcement learning (RL) alignment techniques, such as DeepSeek-R1 \cite{guo2025deepseekr1}. These online RL alignment techniques typically require substantial training resources, high-quality datasets, and complex reward environments, making their application highly resource-intensive.
In contrast, offline alignment methods such as Focused-DPO approximate similar optimization objectives while introducing necessary simplifications and derivations. 
This allows Focused-DPO to achieve comparable or even equivalent optimization results with significantly lower resource requirements. 
Moreover, we leverage prior knowledge discovered in this work: the insight that only a small part of the generated code—specifically, the Error-Prone Points—plays a critical role in determining the overall correctness of the output. By incorporating this insight into the training loss design, we further enhance training efficiency and effectiveness.
Focused-DPO’s low resource requirements and reliable performance make it applicable to a wide range of code generation scenarios. Further exploration of how Focused-DPO compares to these advanced RL techniques in performance and efficiency remains an area for future investigation.

\paragraph{Dataset Construction Strategy}
In Focused-DPO, we introduce a dataset construction technique named \textbf{Error-Prone Identification}, which automatically identifies error-prone points in generated code. 
The primary focus of this paper is on error-prone points associated with correctness in the final output code. However, other factors in source code, such as efficiency, readability, and security, are equally important for optimization.
Exploring whether these factors also reveal "Error-Prone Points" in source code is an intriguing direction for future work. For example, techniques like static code analysis, code smells detection, and identification of common vulnerabilities could help identify and penalize insecure patterns during data construction, leading to safer and more robust code generation.

Additionally, our dataset construction pipeline includes specific design choices, such as the use of a page-rank mechanism and the identification of error-prone points based on common prefixes and suffixes. 
Our preliminary experiments suggest that these settings effectively support the performance of Focused-DPO. Detailed discussions on these designs are provided in Section \ref{sec:discussion}.

\bibliography{main}

\clearpage
\newpage
\appendix

\section{Data Scaling For Focused-DPO}

In our experiments, we use the policy model to sample the dataset for training, with the dataset statistics provided in Table \ref{tab:dataset-statistics}. We also explore how scaling the training data affects the final performance of Focused-DPO. Specifically, we investigate two additional settings: doubling the original training dataset to 10k samples and halving the dataset to 2.5k samples, to observe how these changes impact the effectiveness of the model after Focused-DPO training. 
The experimental results are presented in Table \ref{tab:datascaling}. The results indicate that fine-grained preference optimization converges efficiently within our given data range, and increasing the dataset size does not significantly improve the results.

\begin{table}[h!]
    \centering
    \resizebox{\linewidth}{!}{
    \begin{tabular}{lcc}
        \toprule
        \textbf{Data Scaling} & \textbf{HumanEval / HumanEval+} & \textbf{MBPP / MBPP+} \\
        \midrule
        Qwen2.5-coder-instruct-7B & 0.915 / 0.841 & 0.828 / 0.714 \\
        \midrule
        Focused-DPO (5k) & 0.926 / 0.878 & 0.846 / 0.761 \\
        \midrule
        Decrease to 2.5k & 0.926 / 0.847 & 0.830 / 0.719 \\
        Increase to 10k & 0.926 / 0.878 & 0.843 / 0.756 \\
        \bottomrule
    \end{tabular}
    }
    \caption{Dataset Scaling for Focused-DPO based on \textit{Qwen2.5-Coder-Instruct-7B}}
    \label{tab:datascaling}
\end{table}

\section{Dataset Statistics}

\begin{table}[h]
    \centering
    
    \resizebox{\linewidth}{!}{
    \begin{tabular}{lccc}
        \toprule
        \textbf{Dataset}            & \multicolumn{2}{c}{\textbf{Problems}} & \textbf{Avg. Hidden Tests} \\
        \midrule
        HumanEval                   & \multicolumn{2}{c}{\multirow{2}{*}{164}} & 9.57 \\
        HumanEval+                  & \multicolumn{2}{c}{}                      & 748.07 \\
        MBPP                        & \multicolumn{2}{c}{\multirow{2}{*}{378}} & 3.11   \\
        MBPP+                       & \multicolumn{2}{c}{}                      & 105.40 \\
        \multirow{3}{*}{LiveCodeBench} & Easy  & 279        & 18.07 \\
                                       & Medium   & 331     & 21.81 \\
                                       & Hard  & 270        & 24.78 \\
        \bottomrule
    \end{tabular}
    }
    \caption{Statistics of Evaluation Benchmark.}
    \label{tab:dataset-statistics}
\end{table}

\begin{table}[h]
    \centering
    \resizebox{0.8\linewidth}{!}{
    \scriptsize
    \begin{tabular}{lc}
        \toprule
        \textbf{Statistics based on Qwen2.5-Coder-Instruct-7B}             &    \\
        \midrule
        \multicolumn{1}{l}{\textit{Problems}} & \\
        \midrule
        Training Set                  & 5000              \\
        Validation Set                & 1000              \\
        \midrule
        \multicolumn{1}{l}{\textit{Average Token Lengths}} & \\
        \midrule
        Common Prefix                 & 78.17      \\
        Common Suffix                 & 33.98       \\
        Chosen Mid                    & 57.37      \\
        \hspace{2em} of Total Chosen Code  & 34\%              \\
        Rejected Mid                  & 42.63       \\
        \hspace{2em} of Total Rejected Code  & 28\%              \\
        \bottomrule
    \end{tabular}
    }
    \caption{Training Dataset Statistics based on Qwen2.5-Coder-Instruct-7B}
    \label{tab:training-dataset-statistics}
\end{table}

\section{Scalability and Dataset Independence}

We aim for our proposed method to be scalable and minimally dependent on external prompt resources. Inspired by the \texttt{self-OSS} methodology, we envision that both question generation and solution alignment can be achieved through self-alignment and self-improvement using the policy model itself---a key step toward our ultimate goal.

On the other hand, our method is intentionally designed to avoid being constrained by specific datasets, unlike approaches such as Code-Optimise. Our aspiration is to develop a stable and robust pathway for models to continually enhance their capabilities autonomously.

In our paper, Focused-DPO achieves improvements using preference feedback directly from the model itself. This avoids potential risks associated with externally distilled data, such as unintentionally transferring knowledge from a more capable model to a less capable one.

To further validate the effectiveness of Focused-DPO, we plan to extend our evaluation to existing code datasets. We have adopted the sample datasets from CodeContests. We use the provided test cases to verify the solutions and filter out those samples that have no correct solutions. The final sample number in this dataset setting is nearly 4500. The experiment results are shown in the following Table \ref{tab:codecontesttrain}.

\begin{table}[h]
\centering
\resizebox{\linewidth}{!}{
\begin{tabular}{lccccc}
\toprule
Model & HumanEval & HumanEval+ & MBPP & MBPP+ & LiveCodeBench \\
\midrule
Qwen2.5-coder-instruct-7B & 0.915 & 0.841 & 0.828 & 0.714 & 0.312 \\
+ CodeContests dataset & & & & & \\
Focused-DPO & 0.920 & 0.860 & 0.844 & 0.749 & 0.334 \\
DPO & 0.914 & 0.847 & 0.833 & 0.725 & 0.308 \\
\bottomrule
\end{tabular}
}
\caption{Focused-DPO and DPO Results on CodeContests-train}
\label{tab:codecontesttrain}
\end{table}

These results demonstrate that Focused-DPO is both dataset-agnostic and scalable. We plan to expand testing across additional datasets like APPs in future work to further validate reproducibility and fair comparisons.

\section{Evaluation of $k$'s Impact}

We conducted experiments to measure the consistency of error-prone regions identified across sampled code sequences for varying $k$ values, as shown in Figure \ref{tab:phik}.
The Phi coefficient results reveal that moderate $k$ values (e.g., $k=10$) provide the most consistent identification of error-prone areas.

\begin{table}[h]
\centering

\resizebox{\linewidth}{!}{
\begin{tabular}{cccccc}
\toprule
$k$ Value & Strategy & Correct & Incorrect & Phi Coefficient \\
\midrule
$k = 5$ & Common Prefix + Chosen Mid & 0.5926 & 0.0935 & 0.5237 \\
& Common Prefix + Reject Mid & 0.0132 & 0.4629 & -0.5314 \\
\midrule
$k = 10$ & Common Prefix + Chosen Mid & 0.6367 & 0.0911 & 0.5651 \\
& Common Prefix + Reject Mid & 0.0116 & 0.5575 & -0.6085 \\
\midrule
$k = 20$ & Common Prefix + Chosen Mid & 0.6521 & 0.0975 & 0.5710 \\
& Common Prefix + Reject Mid & 0.0102 & 0.5473 & -0.6025 \\
\bottomrule
\end{tabular}
}
\caption{Phi Coefficient for different $k$ settings}
\label{tab:phik}
\end{table}


We further validated the impact of $k$ by conducting training experiments with different values. The results in Figure \ref{tab:kfordown} indicate that moderate $k$ values yield stable performance improvements across benchmarks.

\begin{table}[h]
\centering

\resizebox{\linewidth}{!}{
\begin{tabular}{cccccc}
\toprule
$k$ Value & HumanEval & HumanEval+ & MBPP & MBPP+ & LiveCodeBench Avg. \\
\midrule
$k = 5$ & 0.920 & 0.871 & 0.835 & 0.748 & 0.336 \\
$k = 10$ & 0.927 & 0.878 & 0.847 & 0.762 & 0.339 \\
$k = 20$ & 0.927 & 0.878 & 0.838 & 0.751 & 0.339 \\
\bottomrule
\end{tabular}
}
\caption{Benchmark results for different $k$ settings based on Qwen2.5-coder-instruct-7B}
\label{tab:kfordown}
\end{table}

\section{Suffix in Positive Loss}

In the original DPO loss, penalizing inappropriate negative samples tends to significantly harm performance. However, retaining a portion of the positive sample learning component improves the robustness and stability of model training. Several existing works, including \cite{liu2024provably}, have analyzed this phenomenon theoretically.
We have also empirically verified this observation in our preliminary experiments. The results, presented in Table \ref{tab:suffixablation}, demonstrate consistent performance improvements.

\begin{table}[h]
\centering
\resizebox{\linewidth}{!}{
\begin{tabular}{lccccc}
\toprule
Model & HumanEval & HumanEval+ & MBPP & MBPP+ & LiveCodeBench \\
\midrule
Qwen2.5-coder-instruct-7B & 0.915 & 0.841 & 0.828 & 0.714 & 0.312 \\
Focused-DPO & 0.927 & 0.878 & 0.847 & 0.762 & 0.339 \\
No Suffix in Positive Reward & 0.927 & 0.872 & 0.844 & 0.760 & 0.334 \\
\bottomrule
\end{tabular}
}
\caption{Benchmark Results Comparison for ``No Suffix in Positive Reward''}
\label{tab:suffixablation}
\end{table}

\section{Case Studies for Error-Prone Points}

\begin{figure*}[h]
\centering
  \includegraphics[width=2\columnwidth]{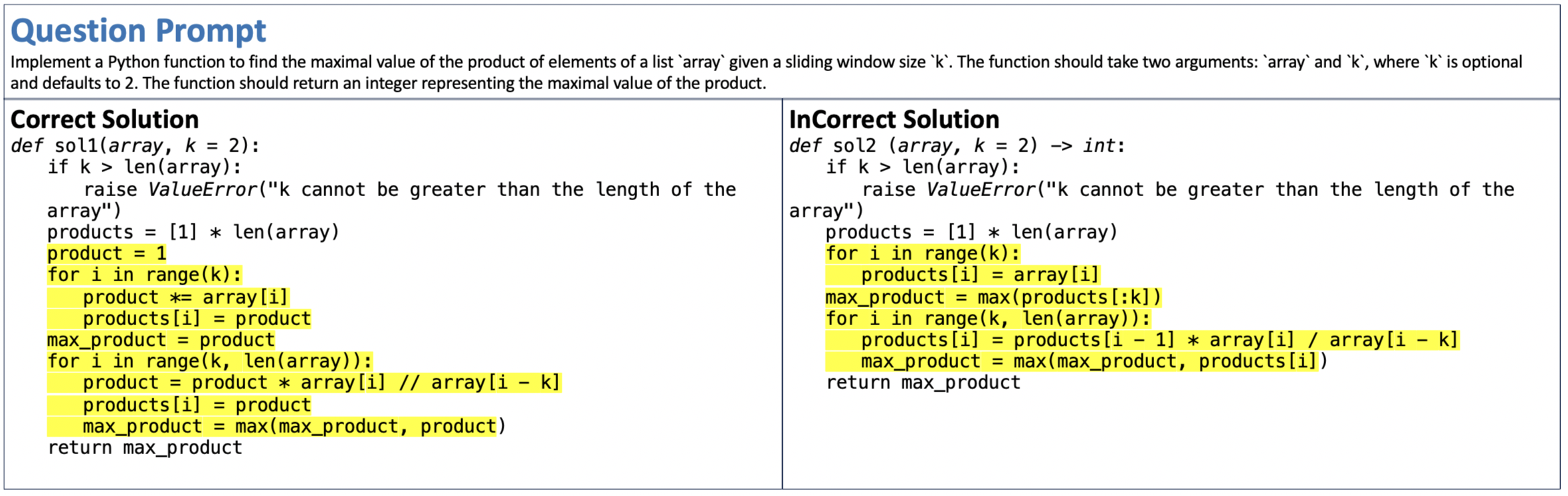}  
\caption{Case Study: max product}
\label{fig:casestudy3}
\end{figure*}

We show some case studies for error-prone points based on \textit{Qwen2.5-Coder-instruct} in the following Figure \ref{fig:casestudy1}, \ref{fig:casestudy2} and \ref{fig:casestudy3}.
\begin{figure*}[h]
\centering
  \includegraphics[width=2\columnwidth]{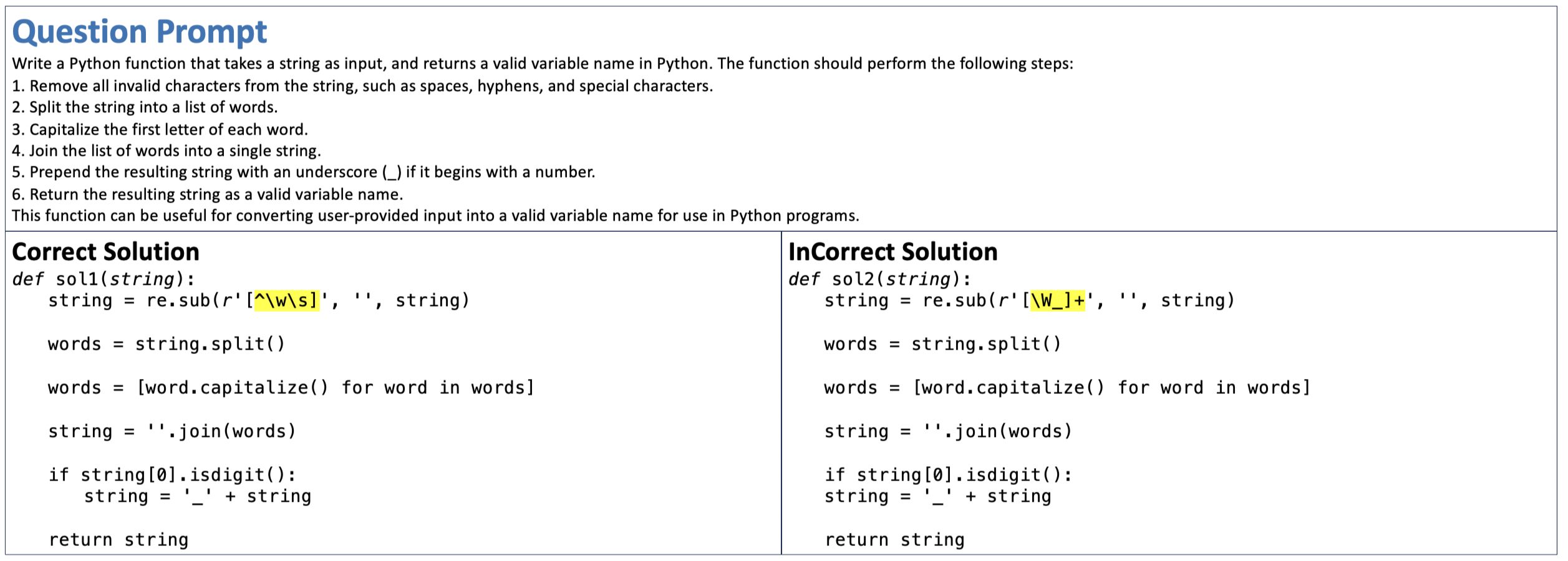}  
\caption{Case Study: convert to valid variable name}
\label{fig:casestudy1}
\end{figure*}
\begin{figure*}[h]
\centering
  \includegraphics[width=2\columnwidth]{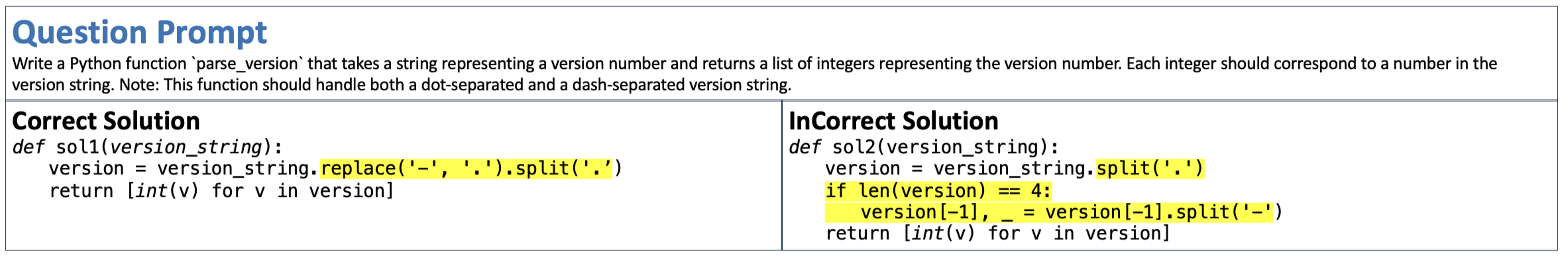}  
\caption{Case Study: parse version}
\label{fig:casestudy2}
\end{figure*}

\end{document}